# Demand- and Priority-Aware Adaptive Congestion Control for Heterogeneous V2X Service Requirements

Miguel Sepulcre, Javier Tortosa-Garcia, Javier Gozalvez
Networked Systems lab, Universidad Miguel Hernández de Elche (UMH), Spain
Email: msepulcre@umh.es, javier.tortosa01@goumh.umh.es, j.gozalvez@umh.es

*Abstract*— Vehicle-to-Everything (V2X) communications enable the exchange of information among vehicles to improve road safety and traffic efficiency. As V2X deployments progress, vehicles are expected to support an increasing number of V2X services, often characterized by different priorities and data transmission requirements. However, existing V2X congestion control mechanisms primarily focus on maintaining channel load stability and fairness at the vehicle level, typically assuming homogeneous traffic demands. This paper proposes a demand- and priority-aware adaptive congestion control technique that explicitly accounts for heterogeneous and time-varying V2X service requirements. The results demonstrate that the proposed technique improves the satisfaction of V2X service demands while maintaining stable channel operation. The proposed technique aligns with current V2X standards, preserving backward compatibility while providing enhancements consistent with ongoing standardization activities.

*Keywords—V2X, V2V, congestion control, Adaptive DCC, resource management, heterogeneous services, CAM, CPM, multi-channel operation, connected vehicles, automated vehicles.*

## I. INTRODUCTION

Vehicle-to-Everything (V2X) communications enable the exchange of information among vehicles with the objective of improving road safety and traffic efficiency. The deployment of V2X-capable vehicles has already begun, and it is expected to accelerate in the coming years. In parallel, an increasing number of V2X services are being defined and progressively introduced into vehicles [1]. As a result, future V2X networks will be characterized by the coexistence of heterogeneous vehicles, not only in terms of capabilities, but also in terms of the number and type of V2X services they support. Even when vehicles implement the same V2X services, their resource requirements may differ significantly. A clear example is the Collective Perception Service (CPS), where vehicles equipped with more advanced sensors are able to detect a larger number of objects, which directly translates into increased data transmission requirements.

Congestion control in V2X networks is a fundamental component to ensure that the shared wireless channel remains stable and usable. Existing congestion control mechanisms, including those standardized to date, have primarily focused on maintaining the channel load within acceptable limits and on ensuring fairness among vehicles [2][3][4]. However, these mechanisms typically assume homogeneous traffic demands and do not explicitly address the challenges introduced by heterogeneous and dynamically varying V2X service requirements. As a consequence, current approaches may fail to efficiently support the coexistence of diverse V2X services and vehicles, particularly when service priorities and asymmetric data transmission demands need to be considered.

This paper addresses these challenges by proposing a Demand- and Priority-Aware (DPA) adaptive congestion control technique that explicitly accounts for heterogeneous and time-varying V2X service requirements. These aspects are currently being actively analyzed within ETSI as part of the evolution towards Release 2 [5], and the proposed technique is designed to align with this evolution by preserving backward compatibility while remaining fully consistent with existing congestion control principles.

## II. CONGESTION CONTROL FOR HETEROGENEOUS V2X SERVICES

Congestion control constitutes a key component of the Resource Management (RM) framework currently being developed by ETSI as part of the evolution towards Release 2 [5]. One of the objectives of RM is to coordinate the use of shared radio resources among multiple V2X services while preserving stable channel operation and compatibility with existing deployments. RM operates across multiple protocol layers, with the congestion control located at the Facilities layer, where V2X service requirements can be observed and translated into transmission constraints. Within this context, congestion control regulates channel utilization by determining the resources available to each vehicle.

### A. Adaptive DCC

ETSI standardized in Release 1 the so called Adaptive Decentralized Congestion Control (DCC) mechanism [4], which is based on the Linear Message Rate Integrated Control (LIMERIC) algorithm [2]. LIMERIC has been extensively analyzed in the literature and has consistently demonstrated stable convergence, fairness, and efficient channel utilization under a wide range of traffic densities and conditions [3]. As a result, it is widely regarded as one of the most effective distributed congestion control solutions for vehicular networks. Its behavior and potential extensions are currently being revisited within the ETSI Release 2 activities [5].

LIMERIC is a decentralized congestion control mechanism that adapts the maximum fraction of channel resources that a vehicle is allowed to use as a function of the observed channel load. Its objective is to drive the overall channel utilization toward a desired target while ensuring fair resource sharing among neighboring vehicles. In this context, fairness corresponds to an equal sharing of the available channel resources among vehicles, independently of the V2X services they implement. To this aim, each vehicle periodically adapts its transmission rate based on the difference between a target channel load and the channel load currently experienced, typically measured through the Channel Busy Ratio (CBR), defined as the fraction of time during which the channel is sensed as busy. The core control equation can be expressed as:

$$\delta(t) = (1 - \alpha) \cdot \delta(t - \Delta T) + \beta \cdot (CBR_T - CBR) \quad (1)$$

where $\delta(t)$ denotes the maximum fraction of time that the vehicle is allowed to transmit at time $t$, $CBR_T$ is the target channel load. The parameters $\alpha$ and $\beta$ jointly determine convergence speed, stability, and the equilibrium operating point of the system, with $\beta$ playing a critical role [2].

Once $\delta$ is determined, the available resources must be distributed among the services implemented by the vehicle. Since Adaptive DCC defines the resources at the vehicle level, a service-level resource distribution mechanism is required in scenarios involving multiple simultaneous V2X services. In this work, resource distribution is performed based on the principle of proportional fairness. Services with the same priority level receive resources in proportion to their required data rates. The distribution follows a tiered process that starts with the highest-priority services; if resources remain, they are subsequently allocated to lower-priority services. When the available resources are insufficient to fully satisfy all services of a given priority level, each service is allocated the same fraction of its requested data rate.

*B. Demand- and Priority-Aware Congestion Control*

To overcome the limitations of Adaptive DCC, where all vehicles are assigned the same number of resources regardless of their actual demands, this paper proposes DPA, a demand- and priority-aware adaptive congestion control technique. The proposed technique builds upon Adaptive DCC and the proportional fairness principle previously described. In addition, it exploits the property that vehicles operating with different $\beta$ values converge to steady-state resource shares ($\delta$) proportional to those $\beta$ values, while preserving stability, convergence [6]. To support time-varying V2X service demands, in DPA the $\beta$ parameter of each vehicle is dynamically adapted according to its V2X service requirements. As a result, vehicles with higher aggregate demands are assigned larger $\beta$ values, whereas vehicles with lower requirements operate with smaller values. The adaptation is performed locally and periodically every time a new CBR measurement is provided by the lower layers, and all vehicles follow the same $\beta$ computation rule to ensure consistent behavior across the network. Backward compatibility is preserved, since legacy Release 1 stations continue to operate with a constant $\beta$ value. A linear adaptive formulation is adopted, where $\beta$ is computed proportionally to the total resources required by the vehicle:

$$\beta = \beta_{\text{base}} \cdot \frac{R_{tot}}{R_{\text{base}}} \quad (2)$$

where $\beta_{\text{base}}$ denotes the $\beta$ value used by Release 1 vehicles employing Adaptive DCC, $R_{\text{base}}$ is a static parameter corresponding to the average resource demand of such stations, and $R_{tot}$ represents the aggregate resource requirements reported by the V2X services implemented by the vehicle that actively generate traffic. Services whose assigned data rate is reduced to zero (typically low-priority services under high channel load conditions) are excluded from the $R_{tot}$ computation. This prevents non-served low-priority services from artificially inflating the requirements of a vehicle. For instance, consider two vehicles that both implement a high-priority service, while only one of them also implements a low-priority service. If the available resources are insufficient to support the low-priority service, the desired behavior is that both vehicles are allocated the same amount of resources for the high-priority service. This is achieved when both vehicles compute identical $\beta$ values and, consequently, obtain the same resource share ($\delta$). By excluding non-transmitting low-priority services from the $\beta$ computation, the proposed technique ensures this behavior.

DPA additionally includes a mechanism to address situations in which service prioritization combined with adaptive $\beta$ may still lead to inconsistent decisions across vehicles. A vehicle may locally infer, based on its computed $\delta$ and service priorities, that it should reduce the data rate of a high-priority service to satisfy the channel load target. However, neighboring vehicles may still transmit lower-priority services, resulting in an inconsistent enforcement of priorities at the network level. In such cases, high-priority traffic would be unnecessarily penalized while less critical transmissions continue consuming resources. To prevent this situation, with DPA, vehicles monitor the messages transmitted by their neighbors and identify the lowest-priority service currently active in the channel. When lower-priority services are detected, the vehicle temporarily overrides the local resource limitation and allows its highest-priority service to transmit at the required data rate. This mechanism ensures that high-priority services are constrained only after lower-priority services have already been limited, enabling consistent priority enforcement across vehicles under resource-limited conditions. Service and message priorities are already defined in V2X standards, and the proposed mechanism ensures their consistent enforcement across vehicles.

III. V2X SCENARIOS AND SERVICES

*A. Single-hop scenario with generic V2X services*

We first evaluate the congestion control mechanisms in a controlled scenario, with stable and well-defined conditions. It consists of a static single-hop topology with 60 vehicles that run three different generic V2X services defined by the 3GPP guidelines [7] with constant requirements. Three vehicle types are defined to represent heterogeneous data rate requirements. Vehicles of Type 1 implement only V2X Service 1, vehicles of Type 2 implement V2X Services 1 and 2, and vehicles of Type 3 implement V2X Services 1, 2, and 3. V2X Service 1 models low-load periodic traffic generating messages every 100 ms according to a predefined sequence of packet sizes {300, 190, 190, 190, 190} bytes. This configuration results in a low average required data rate of approximately 17 kbps. V2X Service 2 corresponds to a high-load periodic traffic service. In this case, packets are generated every 10 ms. Packet sizes follow a discrete distribution, taking values of 1200 bytes with probability 0.2 and 800 bytes with probability 0.8. Under these conditions, the average required data rate is approximately 272 kbps. Finally, V2X Service 3 captures a medium-load aperiodic traffic pattern, introducing randomness in the message generation process. The inter-packet interval is defined as the sum of a fixed 50 ms component and an exponentially distributed random variable with a mean of 50 ms, resulting in irregular transmission times. Packet sizes are uniformly distributed between 200 and 2000 bytes, with a granularity of 200 bytes. The resulting average required data rate is around 56 kbps.

V2X services adapt their message generation according to the decisions provided by the congestion control mechanism,

with each service applying a different adaptation strategy. This design choice allows evaluating the proposed technique independently of the specific adaptation mechanism implemented by the services. Specifically, V2X Service 1 adapts the message generation interval, V2X Service 2 adjusts the size, and V2X Service 3 jointly adapts both parameters.

## B. Highway with CAS and CPS

A second scenario is considered to target a higher degree of realism under conditions closer to those encountered in real deployments. It models a 6-lane highway environment where vehicles move at different speeds and perform realistic driving maneuvers, including lane changes and overtaking. In this scenario, all vehicles implement both the Cooperative Awareness Service (CAS) and the Collective Perception Service (CPS). The CAS handles the generation of Cooperative Awareness Messages (CAMs). Each CAM has a fixed size of 250 bytes and includes the transmitting station's kinematic information, including position, speed, and heading. CAM generation follows dynamic rules that are checked by default every 100 ms. A new CAM is generated whenever the variation of any kinematic parameter (e.g. position, speed or heading) exceeds a predefined threshold, or when a maximum interval of 1 s since the last transmission is reached. In the considered scenarios, the CAS generates a relatively stable load, with data transmission requirements ranging between 6.2 and 7 kbps. To comply with the limits imposed by congestion control, CAS performs message rate adaptation by adjusting the interval at which the CAM generation rules are evaluated.

The CPS is responsible for generating Collective Perception Messages (CPMs), which primarily report the position, velocity, and type of objects detected by the vehicle's onboard sensors. In line with the latest draft of the ETSI technical specification on collective perception, each detected object is assigned a Value of Information (VoI). To concentrate on the resource management aspects, the VoI is here defined as a normalized value between 0 and 1, inversely proportional to the distance between the detecting vehicle and the object as:

$$voi(d) = \max(1 - d/d_{max}, 0) \quad (3)$$

where $d_{max}$ is the maximum perception distance beyond which the VoI becomes zero. In this study, $d_{max}$ is used to represent different sensor quality levels. To capture the heterogeneity of sensing capabilities found in practice, the vehicle population is divided into two groups. Half of the vehicles are equipped with low-quality sensors, modeled by a maximum perception distance $d_{max} = 50$ m, while the remaining 50% are equipped with high-quality sensors characterized by $d_{max} = 150$ m.

CPMs are generated periodically every 100 ms. By default, CPS includes in each CPM all detected objects with a VoI greater than 0.3 in this study. Each object contributes 60 bytes to the payload, while the CPM header has a size of 35 bytes. To comply with the limits imposed by congestion control, CPS applies content-based adaptation by selecting the detected objects with the highest VoI and including them in the CPM until the allowed transmission limits are met. Vehicles with higher-quality sensors achieve larger VoI values, leading to a higher number of detected objects and, consequently, increased data rate requirements. In our scenario, vehicles with low-quality sensors require between 20 and 30 kbps and vehicles equipped with high-quality sensors require between 50 and 60 kbps.

## IV. EVALUATION WITH GENERIC V2X SERVICES

Simulations were conducted using the Veins-INET Framework over OMNeT++, using ITS-G5 at the lower layers with a default data rate of 6 Mbps. The evaluation considers configurations with identical service priorities and configurations with differentiated service priorities (e.g., V2X Service 1 > V2X Service 2 > V2X Service 3). Every 200 ms, the congestion control algorithm is executed using an updated CBR measurement and refreshed service resource requirements. The CBR is computed over a 200 ms observation window as the sum of the fraction of time during which the channel is sensed as busy (i.e., received signal power above a predefined threshold) and the fraction of time used for local transmissions. In parallel, each vehicle updates the data rate requirements of its generic V2X services based on the messages that were generated, or would have been generated, during the preceding one-second interval. The parameters of the Adaptive DCC are configured following ETSI [4] with a target CBR of 0.68. In the proposed technique, $R_{base} = 17 kbps$ and $\beta_{base} = 0.0012$.

Fig. 1 depicts the temporal evolution of the CBR measured by the vehicles in the considered single-hop scenario. The figure reports the median as well as the 5th, 25th, 75th, and 95th percentiles, and distinguishes between configurations where the generic V2X services are assigned equal or different priorities. The results in Fig. 1 (a) and (b) show that Adaptive DCC achieves the expected stable and convergent behaviour, with limited variability among vehicles. A slightly larger dispersion in the measured CBR values is observed when all services are assigned the same priority, as shown in Fig. 1 (a). This effect is mainly caused by the fact that the message generation interval of V2X Service 1 often exceeds the CBR measurement period. As a result, this service contributes only to a subset of the CBR measurements performed across the network. This issue is currently under discussion within ETSI, and a potential mitigation approach involves the use of adaptive CBR measurement mechanisms. Fig. 1 (c) and (d) demonstrate that stability and convergence are preserved when the proposed DPA technique is applied. Moreover, the system converges closer to the target CBR value compared to Adaptive DCC. This improvement is attributed to the higher $\beta$ values assigned to vehicles with larger traffic demands, which enables a more efficient utilization of the radio channel.

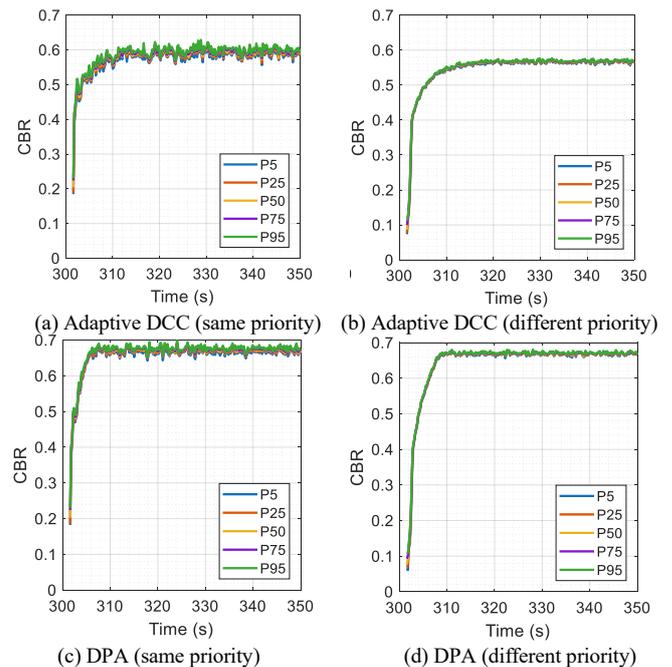

Fig. 1. CBR as a function of time.
(a) Adaptive DCC (same priority)  (b) Adaptive DCC (different priority)
(c) DPA (same priority)  (d) DPA (different priority)

Fig. 2 illustrates the average maximum fraction of resources that a vehicle is allowed to use ($\delta$) and the resulting service satisfaction ratio obtained with Adaptive DCC, distinguishing between configurations where all services are assigned the same priority and those where different priorities are used. As shown in Fig. 2 (a) and (b), all vehicle types converge to the same $\delta$ value, despite having different data rate requirements. This indicates that each vehicle is allocated an identical fraction of the available resources, independently of whether it runs one, two, or three V2X services. This uniform allocation stems directly from the static $\beta$ strategy, which computes $\delta$ solely as a function of the measured CBR and does not incorporate the V2X service requirements. The impact is reflected in the service satisfaction ratios reported in Fig. 2 (c) and (d). When all services are assigned the same priority, as shown in Fig. 2 (c), Type 1 vehicles (with relatively low traffic demands) are able to fully satisfy their service requirements with the allocated resources. In contrast, Type 2 vehicles must proportionally reduce the resources allocated to V2X Services 1 and 2, while Type 3 vehicles experience even more pronounced reductions affecting all three services. These results indicate that, although Adaptive DCC achieves fairness at the vehicle level by enforcing an equal share of resources, it fails to account for the heterogeneous service demands present when multiple services with equal priority coexist. Only when different service priorities are applied, the system exhibits the expected behavior. Fig. 2 (d) shows the satisfaction ratio achieved when the V2X services have different priorities. In this case, the resource demands of V2X Service 1 are always fully satisfied, while no resources are allocated to V2X Service 3. Although Adaptive DCC does not explicitly consider service priorities, it exhibits acceptable behavior in this scenario. This is the case due to the specific traffic configuration rather than an inherent capability of the algorithm. Since Adaptive DCC distributes resources uniformly, vehicles with higher data rate demands are naturally reduced first under congestion, which indirectly favors the lower-rate, higher-priority service. This behavior is largely incidental, as the apparent prioritization results from unused resources of some vehicles being effectively exploited by others, rather than from a correct computation of the resource share ($\delta$) according to service requirements and priorities. For instance, if the highest-priority service were also the one with the largest resource requirements, Adaptive DCC would reduce its transmission rate in the same way as lower-priority services, leading to incorrect prioritization.

Fig. 3 reports the resource allocation and service satisfaction achieved with the proposed DPA technique. When all services are assigned the same priority, Fig. 3 (a) shows that vehicles with higher resource requirements obtain a proportionally larger share of resources thanks to the dynamic computation of $\beta$ based on the service requirements. Fig. 3 (c) indicates that all services experience a similar satisfaction ratio, with their resource demands being reduced by approximately the same proportion with respect to their requested resources (around 21% according to the results). This behavior is expected and represents a clear improvement over Adaptive DCC, as it achieves fairness not only at the vehicle level but at the service level. When different service priorities are applied, the benefits of the proposed DPA technique are maintained. As shown in Fig. 3 (b), vehicles of Type 2 and Type 3 converge to similar $\delta$ values, while vehicles of Type 1 obtain a lower resource share consistent with their reduced requirements. The impact on service satisfaction is illustrated in Fig. 3 (d). In this case, the requirements of V2X Service 1 are fully satisfied for all vehicle types. V2X Service 2 experiences a reduced satisfaction ratio (similar for vehicles of Type 2 and Type 3), while V2X Service 3 is completely suppressed due to its lowest priority and the limited available resources. These results confirm that the proposed DPA technique successfully enforces service prioritization under resource-constrained conditions, while preserving the fairness and efficiency properties. High-priority services are protected from unnecessary degradation, and lower-priority services are progressively limited only when required by channel congestion. Unlike the Adaptive DCC behavior observed, this outcome is not incidental but results from an accurate computation of the resource share ($\delta$) based on the effective service requirements and priorities, which enables consistent prioritization decisions across vehicles.

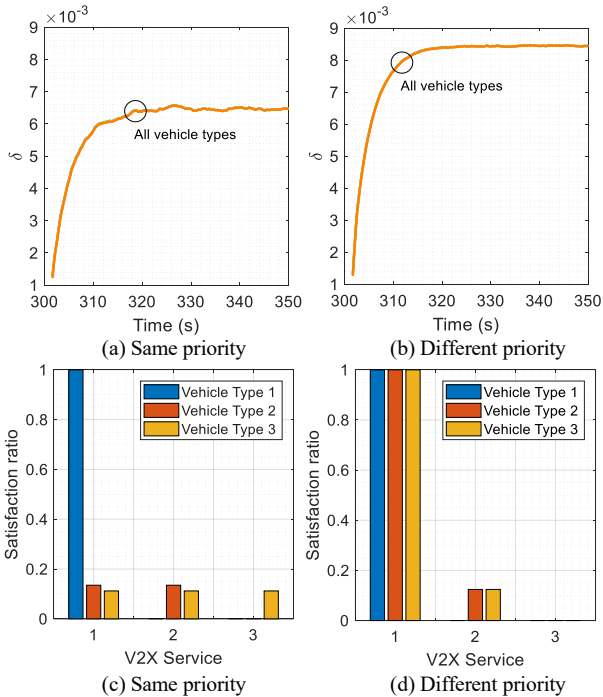

Fig. 2. Resources and service satisfaction ratio for Adaptive DCC.

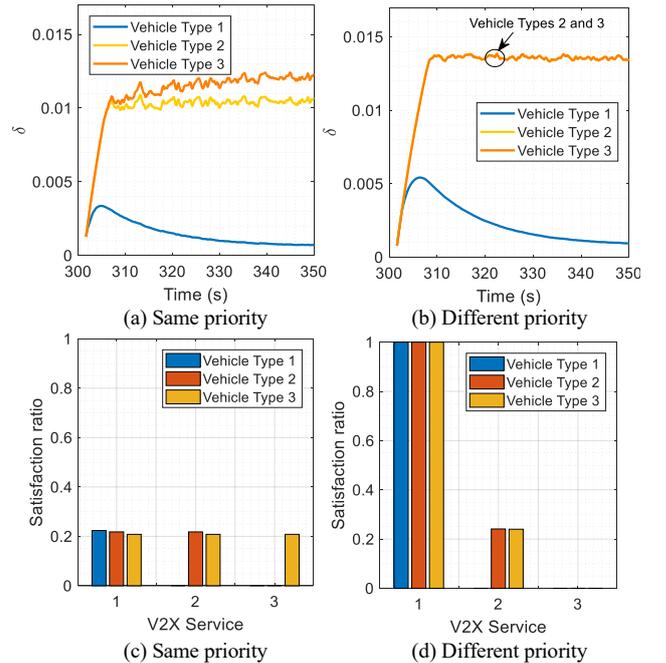

Fig. 3. Resources and service satisfaction ratio for DPA

## V. EVALUATION WITH COOPERATIVE AWARENESS AND COLLECTIVE PERCEPTION SERVICES

The congestion control techniques are also evaluated using the same simulation environment and settings in a highway scenario with all vehicles equipped with CAS and CPS. Fig. 4 presents a boxplot representation of the CBR experienced by all vehicles during the simulation, for both same and different service priority configurations. The boxplots summarize the distribution of the CBR using the 5th, 25th, 50th, 75th, and 95th percentiles, thereby capturing both the central tendency and the variability of the channel load over time and across vehicles. The results show that both Adaptive DCC and DPA achieve a stable CBR and converge close to the target operating point. In particular, the interquartile ranges are narrow in all cases, indicating limited temporal and spatial variability in the experienced CBR. Moreover, the median CBR values remain close to the desired target, especially for the proposed DPA technique. These observations confirm that DPA is able to maintain channel load stability even under heterogeneous and time-varying traffic demands, as introduced by the CAS and CPS services.

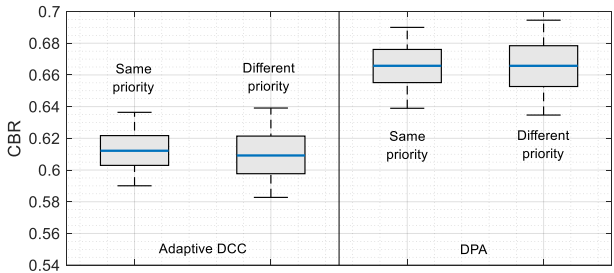

Fig. 4. Boxplot of the CBR experienced by all vehicles during the simulation with the same and different service priority.

Finally, Fig. 5 reports the average service satisfaction ratio obtained in the highway scenario, again differentiating configurations with the same and with different service priorities. When all services are assigned the same priority (Fig. 5 (a)), DPA achieves a similar satisfaction ratio across services and vehicle types. In this case, the available resources are proportionally shared among CAS and CPS, independently of the sensing capabilities of the vehicles. In contrast, Adaptive DCC provides higher satisfaction ratios to vehicles with lower resource requirements (i.e., those equipped with lower-quality sensing capabilities), as resources are uniformly distributed without accounting for heterogeneous demands. This highlights the ability of DPA to accommodate time-varying and heterogeneous requirements more consistently, resulting in a more balanced service satisfaction across the network. When different service priorities are applied (Fig. 5 (b)), Adaptive DCC is able to maintain a high satisfaction level for CAS by reducing the resources assigned to CPS first; however, this comes at the cost of significantly penalizing vehicles equipped with higher-quality sensors. In contrast, DPA successfully preserves the satisfaction of CAS for all vehicles, while the CPS achieves similar satisfaction ratios across vehicles, regardless of sensor quality. These results confirm that DPA effectively enforces service prioritization in realistic scenarios, protecting critical safety-related services while maximizing the utilization of available resources for lower-priority traffic.

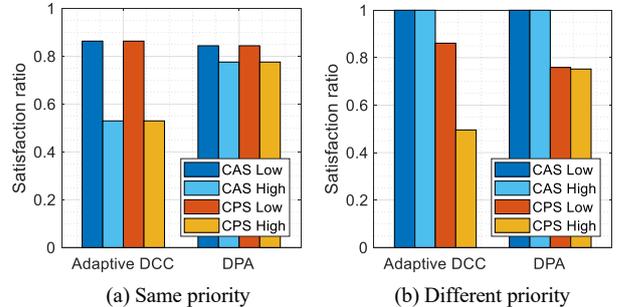

(a) Same priority  (b) Different priority

Fig. 5. Average satisfaction ratio with the same and different service priority. Low and High represent the sensor quality of the vehicle.

## VI. CONCLUSIONS

This paper proposes a demand- and priority-aware adaptive congestion control technique that explicitly accounts for heterogeneous and time-varying V2X service requirements. Extensive evaluations demonstrate that the proposed technique preserves channel load stability and convergence properties comparable to Adaptive DCC while improving the satisfaction of V2X service requirements across vehicles. In particular, it ensures that high-priority services are fully protected under resource-constrained conditions, while lower-priority traffic is adaptively limited when necessary. This behavior is achieved without compromising fairness or stability. The proposed solution remains fully aligned with current V2X standards, preserving backward compatibility while providing important and necessary enhancements consistent with the ongoing ETSI Release 2 evolution.

## ACKNOWLEDGEMENTS

Work partially funded by MICIU/AEI/10.13039/ 501100011033 and "ERFD/EU" (PID2023-150308OB-I00), and Generalitat Valenciana (CIAICO/2024/167).